\documentclass[12pt]{article}
\usepackage{graphicx}
\usepackage{epsfig}
\usepackage{mathtext}
\usepackage[koi8-r]{inputenc}
\usepackage[english]{babel}
\newcommand{\beq}{\begin{equation}}
\newcommand{\eeq}{\end{equation}}
\newcommand{\beqn}{\begin{eqnarray}}
\newcommand{\eeqn}{\end{eqnarray}}
\begin{document} 
 
\title{\textbf{WHAT ARE THE NEUTRINO MASSES}} 
\author{V.P. Efrosinin\\
Institute for Nuclear Research\\
Moscow 117312, Russia}

\date{}
\renewcommand {\baselinestretch} {1.3}

\maketitle
\begin{abstract}
The possible source of the production of neutrino with large masses is
considered. For this purpose the reaction
$\nu_e+n \to e^-+p+\gamma$, in which the electron in $\nu_eeW^+$ vertex is
produced off-mass-shell, is studied.
    
\end{abstract}

In the Standard Model a neutrino does not have any mass. 
Renormalised and gauge invariant mass terms for neutrino might be itroduced
expanding the Model either at the expense of introducing the new sterile states
or total lepton-number violation. In the first case a neutrino is called a
Dirac neutrino, whereas in the latter one it is named as Majorana neutrino.
They are coincided with their antiparticles. The sterile neutrinos do not
couple to the Z and W bosons. The number of the active neutrino states is
restricted by the measurements of the Z boson width and is assumed to be a
three. Following the cosmological data, we can assume the existence of more
than one sterile neutrino mixed with the active ones.  

Neutrino interactions are contained in the CC part of the electroweak theory
Lagrangian:
\begin{eqnarray}
\label{eq:M1}
L^{CC}=-\frac{g}{2\sqrt{2}}j^{CC}_{\alpha}W^{\alpha}+h.c., 
\end{eqnarray}
where $g$ is SU(2) gauge coupling constant and current $j^{CC}_{\alpha}$ reads:
\begin{eqnarray}
\label{eq:M2}
j^{CC}_{\alpha}=2\sum_{l=e,\mu,\tau}\bar{\nu}_e\gamma_{\alpha}l_L. 
\end{eqnarray}
Looking at the Lagrangian (\ref{eq:M1}),(\ref{eq:M2}), we can see that it is
symmetrical relatively to the weak leptonic doublets. 
This symmetry is especially pronounced for the pair $\nu_e,e$, since the
electron mass $m_e$ is highly small and in the calculations at neutrino energy
$\sim 1$ GeV of our interest it is often assumed to be equal zero.

In the present work we consider the following reaction: 
\begin{eqnarray}
\label{eq:M3}
\nu_e+n \to e^{-*}+p, 
\end{eqnarray}
\begin{eqnarray}
\label{eq:M4}
e^{-*} \to e^-+\gamma, 
\end{eqnarray}
that is the reaction
\begin{eqnarray}
\label{eq:M5}
\nu_e+n \to e^-+p+\gamma. 
\end{eqnarray}
We will study the parameters of the reaction (\ref{eq:M3}) with the off-shell
electron in the final state: the mass of this electron, its angular
distribution, cross section etc.
The above parameters are extracted from the analysis of the reaction
(\ref{eq:M5}).
It should be noted that the process (\ref{eq:M3}) and reaction
\begin{eqnarray}
\label{eq:M6}
\nu_e+n \to e^-+p, 
\end{eqnarray}
where $e^-$ is on the mass shell are not coherent.
Accounting for the structureless of the leptons, one can use the Lagrangian
(\ref{eq:M1}),(\ref{eq:M2}) for the description of production of electrons 
both on-mass-shall (\ref{eq:M6}) and off-mass-shell
(\ref{eq:M3}),(\ref{eq:M5}).

It is no doubt, that due to the fact that the Lagrangian
(\ref{eq:M1}),(\ref{eq:M2}) is hermitian
(the CP - violation
$<10^{-3}$ of cross section of weak interactions)
the cross sections of the reactions
\begin{eqnarray}
\label{eq:M6p}
e^-+p \to \nu_e+n 
\end{eqnarray}
and (\ref{eq:M6}) are the same at the equel collision energy 
(assuming that $m_e=0$).

The interaction in the vertex (\ref{eq:M1}),(\ref{eq:M2}) with the off-shell
electron production may be observed only in reaction
(\ref{eq:M5}). The cross section of it is about of 
$\sim \alpha$ from the one of the process (\ref{eq:M6}).
 
If the reaction
\begin{eqnarray}
\label{eq:M7}
e^-+p \to \nu^*_e+n, 
\end{eqnarray}
where $\nu^*_e$ is off-mass-shell, does not exist than CP-violation in
the vertex (\ref{eq:M1}),(\ref{eq:M2}) will be at least at the level of
$\sim \alpha$ on weak interaction. It is possible to say that such reaction is
forbidden due to the uncertainty principle because the life-time of
$\nu^*_e$ is substantially larger than that of
$e^{-*}$. This means large CP-violation in the off-shall production of
$e^{-*}$ and $\nu^*_e$, what is needed to be checked experimentally.
The off-shell spectrum of $\nu^*_e$ from (\ref{eq:M7}) is similar to that of    
$e^{-*}$ in
(\ref{eq:M3}),(\ref{eq:M5}). In the case if the three flavor neutrino mixing
with the sterile neutrino exists, this leads to the transition of the off-mass-
shell neutrino $\nu^*_e$ into the on-mass-shell sterile neutrino
$\nu_s$ with the corresponding mass.
In such a manner the suppressing action of the uncertainty principle is excluded
completely or partly.   

\begin{figure*}[hb]
\begin{center}
\epsfig{file=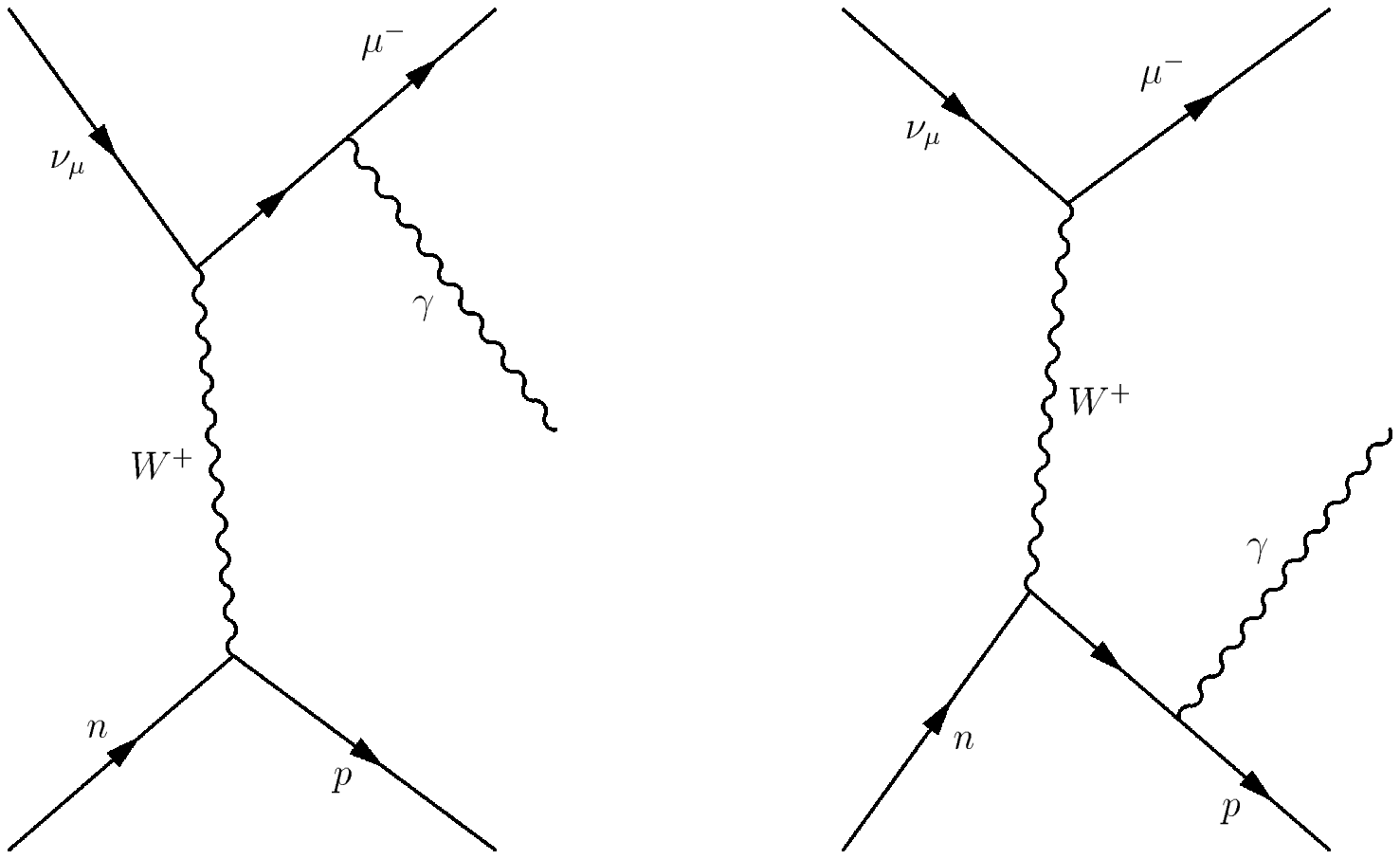,width=14.cm}
\end{center}
\caption{}
\label{fig:fi0}
\end{figure*}

In ref. \cite{efros} the cross section of reaction
\begin{eqnarray}
\label{eq:M8}
\nu_{\mu}+n \to \mu^-+p+\gamma 
\end{eqnarray} 
has been estimated at low neutrino energies ($E_{\nu} \sim 1$ GeV).
This reaction contributes to the one photon background in the experiments on
determination of the neutrino oscillation parameters.
The gauge invariant amplitude of the reaction
(\ref{eq:M8}) has been obtained in \cite{efros}
by adopting the Low theorem with taking into account the two diagram with the
emission of photon, respectively, from muon and proton
(See, Fig.\ref{fig:fi0}).

In the present work we are interested in the off-mass-shell behavior of an
electron. The diagram with the excited electron gives us such possibility.
We are not interested in the exact calculation of the cross section of process
(\ref{eq:M5}) with the employing of the gauge invariant amplitude.
The more so as the second diagram with the photon emission from the proton
contributes to the cross section of reaction
(\ref{eq:M5}) less than the diagram with the exited electron.
We will perform our calculations in order of magnitude of the mass of an
off-shell electron $e^{-*}$.

The cross section of the reaction
\begin{eqnarray}
\label{eq:M9}
\nu_e(k)+n(p) \to e^-(k^{\prime})+p(p^{\prime})+\gamma(r), 
\end{eqnarray}
where in the parenthesis the 4-momenta  of nucleons, leptons and photon are
indicated, in lab system can be represented as follows:
\begin{eqnarray}
\label{eq:M10}
d\sigma&=&\frac{1}{32(2\pi)^4 \omega M}|T|^2 \times \nonumber\\
&\times&\frac{\omega^{\prime}}{|M+\omega(1-cos\theta_{k^{\prime}})
-E_{\gamma}(1-sin\theta_{\gamma} sin\theta_{k^{\prime}}
cos\varphi_{k^{\prime}}-cos\theta_{\gamma} cos\theta_{k^{\prime}})|}
\times \nonumber\\
&\times&E_{\gamma} dE_{\gamma} dcos\theta_{\gamma}
dcos\theta_{k^{\prime}} d\varphi_{k^{\prime}}. 
\end{eqnarray}
Here, $\omega$ is the neutrino energy; M is the nucleon mass;
$\omega^{\prime}$ is the electron energy 
(the mass of the on-mass-shell electron is specified as
$m_e$=0); $E_{\gamma},~\theta_{\gamma}$ are the energy and polar angle of the
photon in the lab frame;
$\theta_{k^{\prime}},~\varphi_{k^{\prime}}$ are, respectively, lab polar
and azimuthal angles of the final electron in the ground state. 

The amplitude of the reaction (\ref{eq:M9}) is given by
\begin{eqnarray}
\label{eq:M11}
T&=&\frac{eG_F cos\theta_C}{\sqrt{2}}\bar{u}(p^{\prime})
\bigl[g_V\gamma^{\mu}+g_M \frac{i\sigma^{\mu \lambda} q_{\lambda}}{2M}
-g_A \gamma^{\mu} \gamma_5 \bigr]u(p)\times \nonumber \\
&\times&\frac{1}{2(k^{\prime} r)} \bar{u}(k^{\prime})
[2(k^{\prime} \varepsilon^*)+\hat{\varepsilon}^* \hat{r}]\gamma_{\mu}
(1-\gamma_5)u(k), 
\end{eqnarray}
where $\varepsilon^*_{\lambda}$ is the polarization of photon,
$q=k-k^{\prime}$,
$g_V,g_M$ and $g_A$ are the weak formfactors of the $\nu n$ interaction
with the charged current \cite{pasc}. \cite{zell}.

\begin{figure*}[hb]
\begin{center}
\epsfig{file=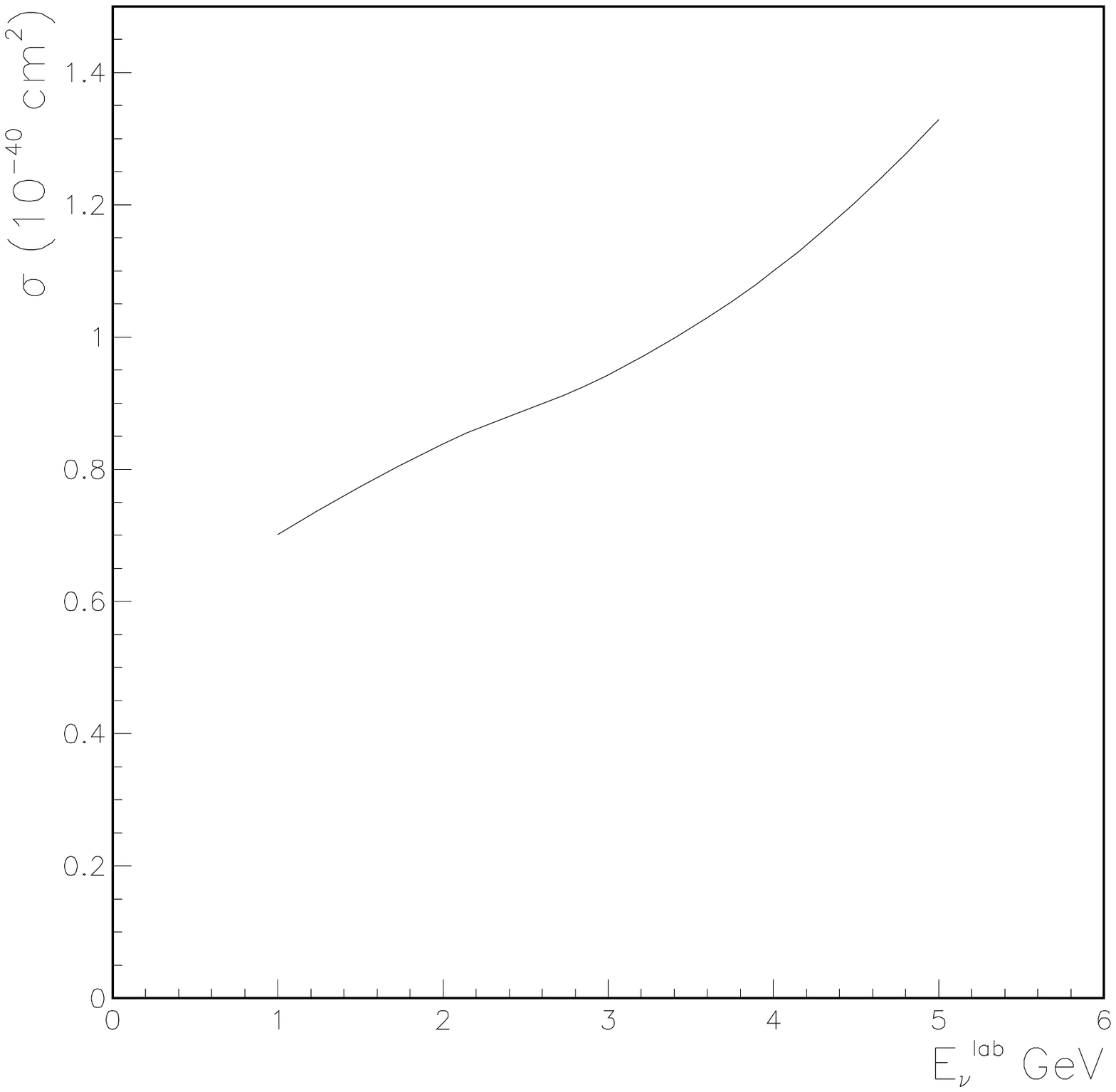,width=14.cm}
\end{center}
\caption{}
\label{fig:fi1}
\end{figure*}

\begin{figure*}[hb]
\begin{center}
\epsfig{file=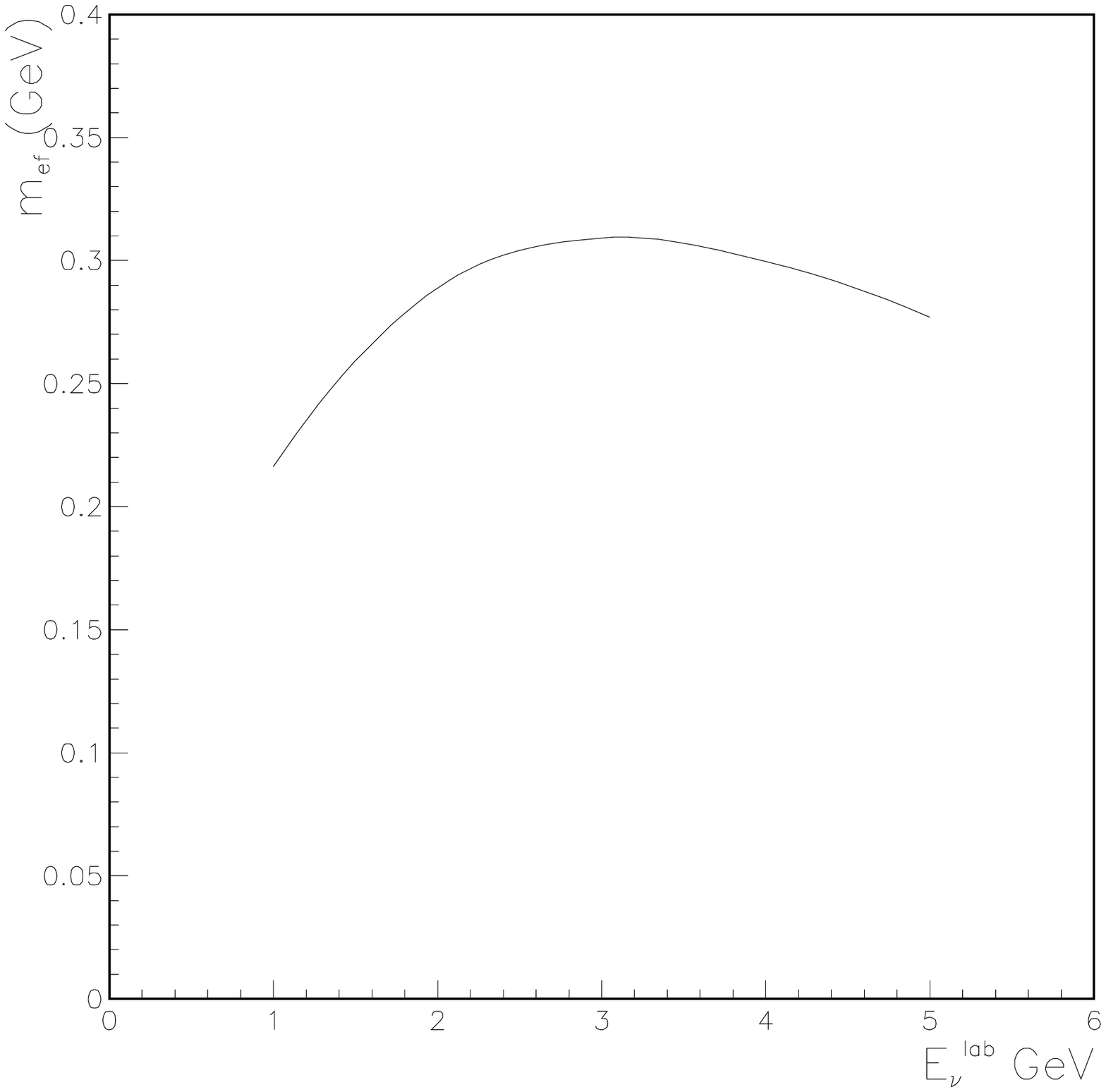,width=14.cm}
\end{center}
\caption{}
\label{fig:fi2}
\end{figure*}

\begin{figure*}[hb]
\begin{center}
\epsfig{file=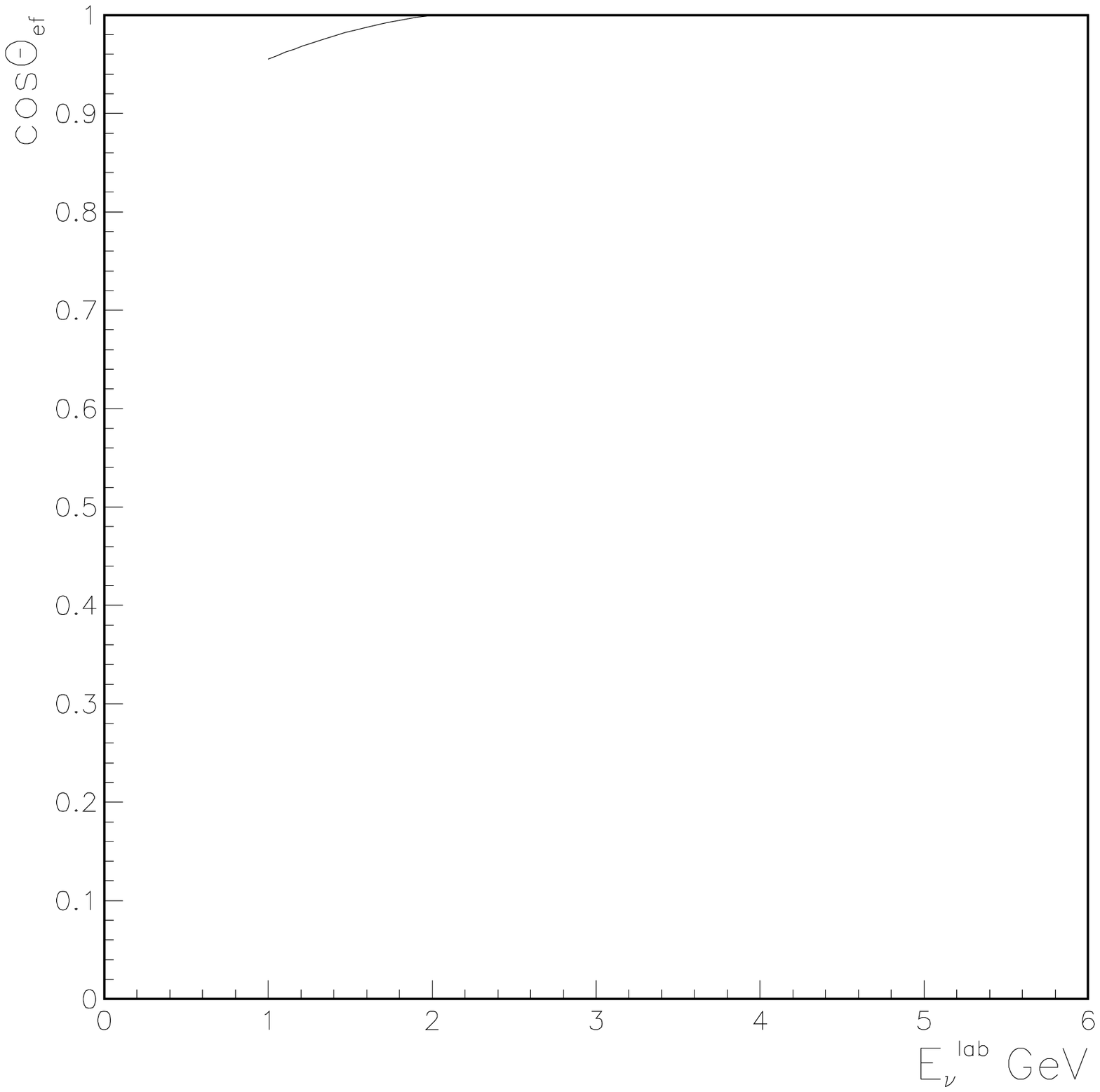,width=14.cm}
\end{center}
\caption{}
\label{fig:fi3}
\end{figure*}

By using eqs. (\ref{eq:M10}), (\ref{eq:M11}),
we calculate in our approach the cross section of the reaction (\ref{eq:M9}).
It is shown in Fig.\ref{fig:fi1} as a function of neutrino energy
$E_{\nu}$ in the region of $1 \div 5$ GeV.
We also calculated the averaged over the cross section the effective masses
(Fig.\ref{fig:fi2}) and polar anglex
(Fig.\ref{fig:fi3}) of the $e^{-*}$.

\begin{figure*}[hb]
\begin{center}
\epsfig{file=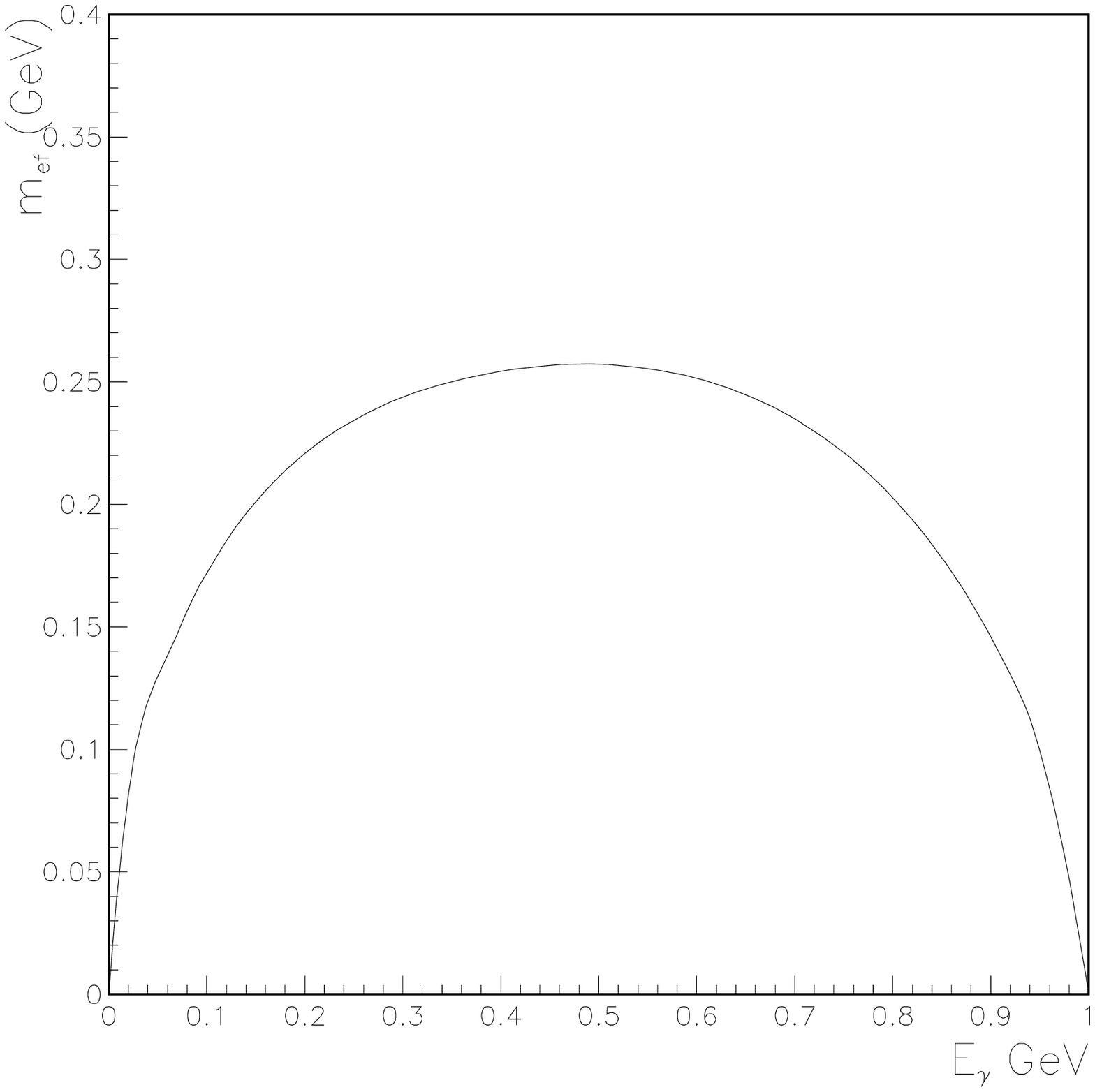,width=14.cm}
\end{center}
\caption{}
\label{fig:fi4}
\end{figure*}

\begin{figure*}[hb]
\begin{center}
\epsfig{file=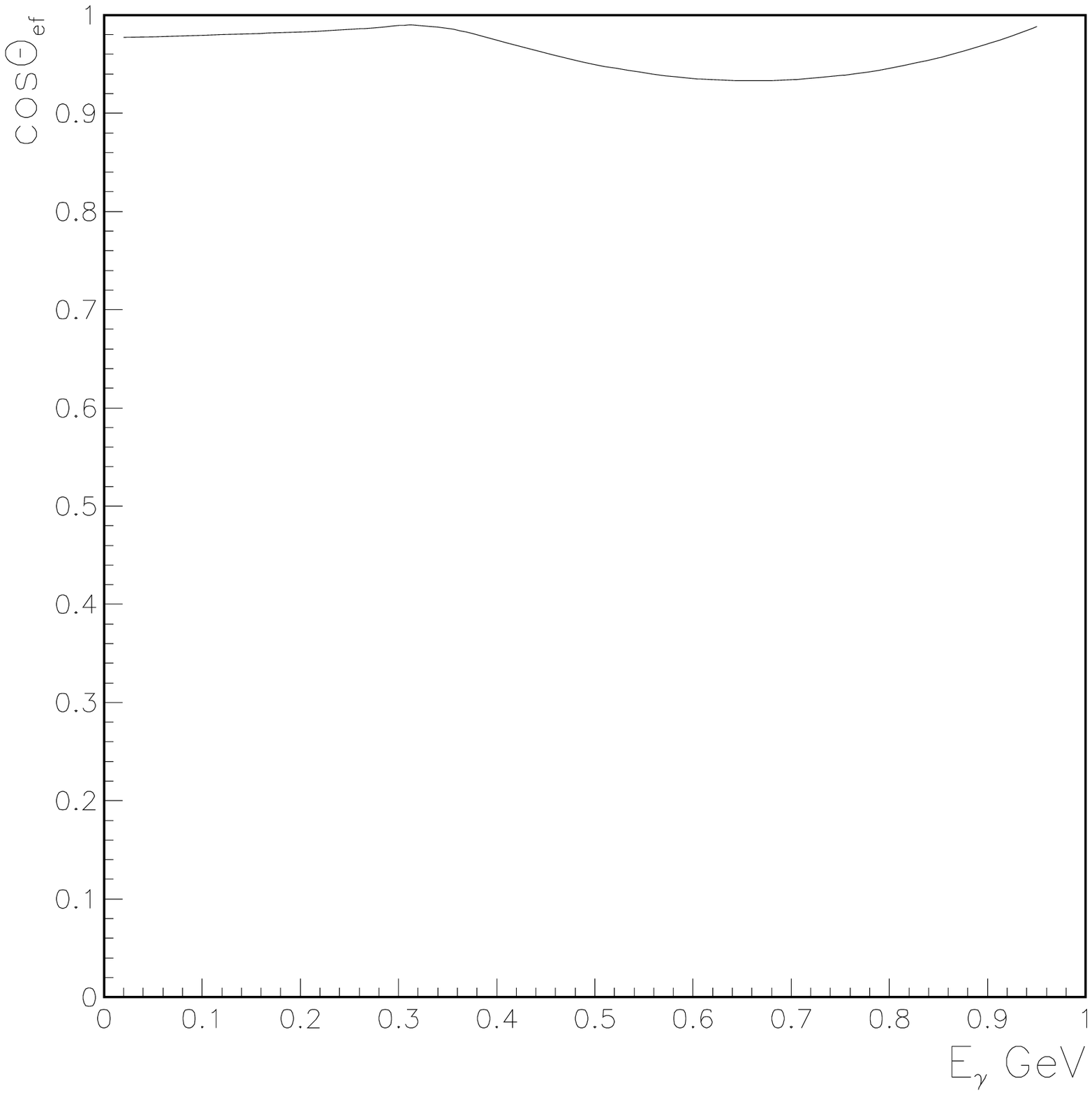,width=14.cm}
\end{center}
\caption{}
\label{fig:fi5}
\end{figure*}

Fig.\ref{fig:fi4} shows a distribution of the $e^{-*}$ effective mass as a
function of photon energy
$E_{\gamma}$, whereas Fig.\ref{fig:fi5} displays a distribution of the
emission angle of
$e^{-*}$ as a function of $E_{\gamma}$ at a fixed neutrino energy
$E_{\nu}=1$ GeV.

Turning back to the preceding relatively to the analogy of reactions
(\ref{eq:M3}) and
(\ref{eq:M7}), it is possible to suggest the following. When the neutrino
with small mass is produced in some reaction, in it is possible also the
creation of neutrino with large masses, depending from the kinematics of
reaction or decay, with the cross section which is suppressed by
$\sim \alpha$ compared to that for production of neutrino with small mass
in this process.
An estimate of these large neutrino masses can be obtained from the analysis
which is similar to that just performed by us for the specific reaction.
This is one of the possible mechanisms for the production of sterile neutrinos.
There is reason to search the neutrino with large masses experimentally.

\newpage
\clearpage

\newpage
\clearpage

\newpage
\clearpage

\newpage
\clearpage

\newpage
\clearpage

\newpage
\clearpage

\newpage
\clearpage

\begin{center}
Figure captions
\end{center}

Fig.~1. The diagram of photon bremsstrahlung in the reaction
$\nu_{\mu}+n \to \mu^-+p+\gamma$.

Fig.~2. The dependence of the cross section of the reaction 
$\nu_e+n \to e^-+p+\gamma$ on the initial neutrino energy
$E_{\nu}$.

Fig.~3. The dependence of the $e^{-*}$ effective mass in the reaction
$\nu_e+n \to e^-+p+\gamma$ on the initial neutrino energy
$E_{\nu}$.

Fig.~4. The dependence of the $e^{-*}$ effective emission polar lab angle
in the reaction
$\nu_e+n \to e^-+p+\gamma$ on the initial neutrino energy
$E_{\nu}$.

Fig.~5. The distribution of the $e^{-*}$ effective mass as a function of
photon energy
$E_{\gamma}$ at a fixed neutrino energy $E_{\nu}=1$ GeV.

Fig.~6. The distribution of the emission polar angle of
$e^{-*}$ as a function of photon energy
$E_{\gamma}$ at a fixed neutrino energy $E_{\nu}=1$ GeV.


\end{document}